\documentclass{article}

\usepackage{PRIMEarxiv}

\usepackage[utf8]{inputenc} 
\usepackage[T1]{fontenc}    
\usepackage[bookmarks=true,bookmarksopen=false,bookmarksnumbered=true,colorlinks=true]{hyperref}
\hypersetup{pagebackref=true,colorlinks=true,citecolor=blue,urlcolor=cyan,linkcolor=red,%
  pdftitle={A model of semantic completion in generative episodic memory},%
  pdfauthor={Zahra Fayyaz, Aya Altamimi, Sen Cheng, Laurenz Wiskott}}
\usepackage{url}            
\usepackage{booktabs}       
\usepackage{amsfonts,amsmath}       
\usepackage{nicefrac}       
\usepackage{microtype}      
\usepackage{lipsum}
\usepackage{fancyhdr}       
\usepackage{graphicx}       
\graphicspath{{media/}}     

\usepackage{microtype}
\DisableLigatures[f]{encoding = *, family = * }

\usepackage[table]{xcolor}
\usepackage{graphicx}
\usepackage{tikz}
\usepackage{subcaption}

\usepackage{array}
\usepackage[color=green!40]{todonotes}
\usepackage{marginnote}
\newcolumntype{+}{!{\vrule width 2pt}}

\newlength\savedwidth

\raggedright
\usepackage[aboveskip=1pt,labelfont=bf,labelsep=period,justification=raggedright,singlelinecheck=off]{caption}

\makeatletter
\renewcommand{\@biblabel}[1]{\quad#1.}
\makeatother

\usepackage{lastpage,fancyhdr,graphicx}
\usepackage{epstopdf}
\fancyheadoffset[L]{2.25in}
\fancyfootoffset[L]{2.25in}
\newcommand{\ignore}[1]{}
  
\title{A model of semantic completion \\ in generative episodic memory}

\author{
    Zahra Fayyaz\textsuperscript{1},
    Aya Altamimi\textsuperscript{1},
    Sen Cheng\textsuperscript{1},
    Laurenz Wiskott\textsuperscript{1*} \\
    \texttt{\textbf{1} Institute for Neural Computation, Ruhr University Bochum, Germany } \\
   \texttt{* laurenz.wiskott@rub.de} 
}

\begin{document}

\maketitle

\begin{abstract}

  Many different studies have suggested that episodic memory is a generative process, but
  most computational models adopt a storage view. In this work, we propose a
  computational model for generative episodic memory.  It is based on the central
  hypothesis that the hippocampus stores and retrieves selected aspects of an episode as
  a memory trace, which is necessarily incomplete.  At recall, the neocortex reasonably
  fills in the missing information based on general semantic information in a process we
  call semantic completion.

  As episodes we use images of digits (MNIST) augmented by different backgrounds
  representing context. Our model is based on a VQ-VAE which generates a compressed
  latent representation in form of an index matrix, which still has some spatial
  resolution. We assume that attention selects some part of the index matrix while others
  are discarded, this then represents the gist of the episode and is stored as a memory
  trace.  At recall the missing parts are filled in by a PixelCNN, modeling semantic
  completion, and the completed index matrix is then decoded into a full image by the
  VQ-VAE.

  The model is able to complete missing parts of a memory trace in a semantically
  plausible way up to the point where it can generate plausible images from scratch.  Due
  to the combinatorics in the index matrix, the model generalizes well to images not
  trained on. Compression as well as semantic completion contribute to a strong reduction
  in memory requirements and robustness to noise. Finally we also model an episodic
  memory experiment and can reproduce that semantically congruent contexts are always
  recalled better than incongruent ones, high attention levels improve memory accuracy in
  both cases, and contexts that are not remembered correctly are more often remembered
  semantically congruently than completely wrong.

\end{abstract}


\reversemarginpar

\section*{Introduction}

Episodic memory is hippocampus dependent and enables us to remember personally
experienced events \cite{clayton2007episodic}, semantic information on the other hand is
represented in higher cortical areas and captures general facts and regularities of the
world around us \cite{reisberg2013oxford}.  Early concepts of episodic memory were based
on the storage model, according to which the content of the memory more or less
faithfully reflects the content of the experience \cite{tulving1972organization}. This
view is oversimplified since it reduces episodic recall to a mere readout process of
stored complete information. However, overwhelming empirical evidence suggests that the
recalled memories can be influenced both by past and future information as well as the
context of encoding and recalling.  Pioneering studies have suggested that semantic
interpretations, rather than sensory inputs, are stored in memory
\cite{bartlett1995remembering,sachs1967recognition} and that memories are reconstructed
during recall \cite{bartlett1995remembering}. In word list studies using the
Deese–Roediger–McDermott (DRM) paradigm \cite{deese1959prediction, roediger1995creating}
participants "remember" semantically related words that were not on the study list when
asked to retrieve the words studied earlier. There is also evidence that semantic and
episodic memories interact and complement each other during retrieval
\cite{greenberg2010interdependence}. For instance, Devitt \textit{et al.}
\cite{devitt2017episodic} have found in a meta-analysis of eight studies based on
autobiographical interviews that when subjects report an episode, their usage of
"internal" (episodic) details and "external" (semantic) details are negatively
correlated. The subjects use semantic information to compensate for insufficient episodic
detail in their memory.  Other examples are experiments by Barlett
\cite{bartlett1995remembering}, where subjects of non-matching cultural background
recalled folk tales. The recalled stories were distorted to match the subjects' cultural
background (semantic information). Finally, There are also paradigmatic examples of
memory adjustment due to social context\cite{hirst2012remembering, deuker2013playing},
self-model \cite{axmacher2010natural}, stress\cite{herten2017role}, and many other
factors \cite{Schacter2020MemoryAI,addis2020mental}.

Few contemporary researchers would oppose the idea that episodic memory is – at least to
a certain degree – generative in nature \cite{greenberg2010interdependence}. Nonetheless,
most of the existing computational models (including some of our own
\cite{cheng2016episodic, cheng2011structure, neher2015memory}) adopt the storage view,
where memories are preserved and later retrieved faithfully \cite{rolls1995model,
  jensen1996novel, becker2005computational}.  Such models are usually tested with either
random patterns or abstract spatial representations \cite{rolls1995model,
  jensen1996novel, becker2005computational, cheng2011structure, neher2015memory} but not
with realistic sensory input. With such artificial input patterns, it is rather
suggestive to think in terms of mere storage memory, since there is not much structure in
the input that could be exploited.  However, there is a rich hierarchical structure of
features and statistical relationships in natural stimuli, which was not considered or
even exploited in these models.

In order to model the generative process of episodic memory recall we believe it is
important (i)~to use (real-world) input patterns as stimuli with enough structure that
can be exploited by a semantic system for a generative process, (ii)~to discard some of
the input patterns during storage to model the inevitable loss of information in the
brain due to the attentional bottleneck, and (iii)~to include a generative element in the
model that is able to reasonably fill in the missing information.  Discarding some of the
input patterns can be done in at least two ways, by lossy compression and by selection
(either before or after compression). The former refers to a process like mp3 encoding, a
compression that tries to discard only the information that is irrelevant or recoverable
from what is being stored. The latter refers to a process where some part is selected for
storage and another one is discarded altogether, e.g.\ from a picture of a water mill at
a creek the mill could be attended to and stored while the creek could be ignored.  When
recalling the mill, our semantic system probably complements it by a creek, but the creek
might look very different from the original one, and we are probably not even aware of
this, a process we refer to as scenario construction \cite{cheng2016dissociating}.  When
recalling the mp3 encoded song, on the other hand, there might be some noise due to the
strong compression, but all in all the song is faithfully reconstructed.  We refer to the
representation reduced by compression and selection as the gist that is stored in a
memory trace and from which the original episode can be reconstructed, either quite
faithfully if only compression is involved or at least plausibly if also selection is
involved.

Following the lossy compression approach from a perspective of rate-distortion theory
(RDT) and efficient coding, Bates and Jacobs \cite{bates2020efficient} as well as Nagy et
al.\ \cite{nagy2020optimal} have modeled perception and episodic memory as a generative
process. Bates and Jacobs have argued that a capacity-limited perceptual system like the
brain should use prior knowledge and take into account task dependencies to compress the
input into an optimal representation. Nagy et al.\ have demonstrated that systematic
distortions in memory are similar to the distortions that are characteristic of a
capacity limited generative model adapted to an environment for compression. They use a
Variational AutoEncoder (VAE) as a model for memory \cite{kingma2013vae}.  A VAE is an
autoencoder architecture that maps input images to a plain Gaussian model distribution
and back again to images.  Episodic memory can then be modeled by storing the location in
the Gaussian as a memory trace, which is a low-dimensional feature vector.  From such a
memory trace the full image can be reconstructed. The lossy compression from input to
Gaussian is so extreme in this case that memory recall is largely generative. It is even
possible to generate new images without any input or memory trace by sampling from the
Gaussian and then decoding this vector.  Any such image looks similar to the images seen
during training, in fact the system can only represent such seen images or interpolations
thereof. These models are already generative but with a focus on optimal compression
while our focus is semantic completion.

The model we propose here is built on a Vector-Quantized Variational AutoEncoder (VQ-VAE)
\cite{oord2017vqvae}, which, despite the similarity of the names, works quite differently
from a VAE. It transforms input images into an array of low-dimensional feature vectors
(rather than just one) and thereby maintains some spatial resolution. This has two great
advantages: (i)~We cannot only model compression but also spatial selection by
attention. We do that by discarding some fraction of the feature vectors in the array and
keeping the rest. (ii)~The model can also store and recall input patterns that are quite
different from those seen during training because the known feature vectors can be
combined in many different new spatial constellations. For instance, a VQ-VAE trained on
digits 0~to~4 can also represent digits 5~to~9, a VAE cannot do that.  We see another
advantage of the VQ-VAE for our purposes in that the feature vectors are quantized, which
is in analogy to semantic categorization in the brain.

However, the VQ-VAE is not generative.  Thus, we add a Pixel Convolutional Neural Network
(PixelCNN), which is able to fill in missing feature vectors in the array of vectors, up
to the point that it creates entirely new arrays from scratch. Conceptually this model is
able to fill in a creek to complement the mill, but we use much simpler stimuli (namely
handwritten digits from the MNIST database) for our simulations for computational
efficiency.

The VAE, the VQ-VAE, and the PixelCNN in the models above are trained on large sets of
images and thereby capture the statistical regularities in them, which we consider to be
semantic information. The VAE and VQ-VAE capture the semantic information needed to do
and undo the compression; the PixelCNN captures the semantic information required to fill
in neglected parts. The term \emph{semantic} might seem overly ambitious here, but we
believe that the semantic information these generative models capture shares essential
characteristics with what we would normally refer to as semantic, namely general
regularities of the world that hold beyond and are represented independently of
particular episodes. Furthermore, in the VQ-VAE the semantic information is quantized,
i.e., \ discrete in nature, which resembles categorization.

One might think that a storage memory would be advantageous over a generative one because
of its faithfulness. However, scenario construction during recall is essential to the
etiological function of episodic memory, because it provides far more flexibility to deal
with missing data and to adjust to variable demands and constraints than a faithful
reproduction of past experiences could. Moreover, the already acquired semantic knowledge
can help to improve the storage efficiency of the memory. Put simply, generativity is a
useful feature in episodic memory, not an aberration\cite{SCHACTER2011467}.

\section*{Computational model of generative episodic memory}

Based on a biologically-motivated conceptual framework and using methods from machine
learning, we have developed a computational model that allows us to investigate semantic
completion in a generative episodic memory on real world images on a fairly abstract
level but still in analogy to concrete brain structures, so that predictions on a
behavioral level but also about neural processes are possible.

\subsection*{Conceptual framework}

We hypothesize that generative episodic memory works as follows:
\begin{enumerate}
\item Sensory input patterns that make up the episode are perceived by a hierarchically
  organized network and transformed into a hierarchical perceptual-semantic
  representation in cortical areas, such as the visual system.
\item Some elements of this representation are selected for storage in episodic
  memory. We call this the episodic gist.
\item The episodic gist is stored in hippocampal memory as pointers to the corresponding
  perceptual-semantic elements in cortical areas. We call this the memory trace.
\item Triggered by some external or internal cue, or even spontaneously, the memory trace
  can be reactivated.
\item The pointers in the memory trace reactivate corresponding perceptual-semantic
  elements in cortical areas.
\item Semantic information in the cortical areas complement the reactivated elements by
  means of a recurrent dynamics to construct a plausible full representation from the
  incomplete gist stored in the memory trace.
\end{enumerate}
Several of these steps and concepts deserve closer consideration.
\begin{itemize}
\item We speak of a \emph{perceptual-semantic} representation, because (i)~we consider
  the transformation from the raw input to a high-level semantic representation a gradual
  process, as is well known for deep neural networks, and (ii)~we believe that we can
  actually remember also quite low level aspects of an episode, such as the exact color
  and shape of an object. So, we believe that there is no clear cut distinction between
  perceptual and semantic representations.
\item The concept of \emph{gist} is well known \cite{koutstaal1997gist, oliva2005gist,
    sachs1967recognition}. The episodic gist \cite{cheng2016episodic} contains essentials
  about the episode that are selected dynamically depending on attention and the context
  \cite{graham2000insights}. They may be detailed in some cases and general and vague in
  others.
\item Episodic \emph{memory traces} are pointers to perceptual-semantic elements of the
  sensory input rather than the representations of the input itself
  \cite{fang2018interaction}.
\item \emph{Semantic information} is usually extracted from multiple experiences, is
  mostly categorical, and refers to the prototypical properties of objects or people and
  their relationships \cite{tulving1972organization, collins1969retrieval}.  Evidence
  from patients with semantic dementia suggests that semantic information is vital for
  episodic memory recall \cite{irish2013pivotal}.
\item Because of its generativ nature, we call the retrieval process \emph{scenario
    construction}.
\end{itemize}
Our computational model does not yet capture all aspects of this conceptual framework,
but we believe it is a good first step in the right direction.

\subsection*{Network architecture} \label{section:Network-arch}

\begin{figure}[htb!]
\includegraphics[width=\textwidth]{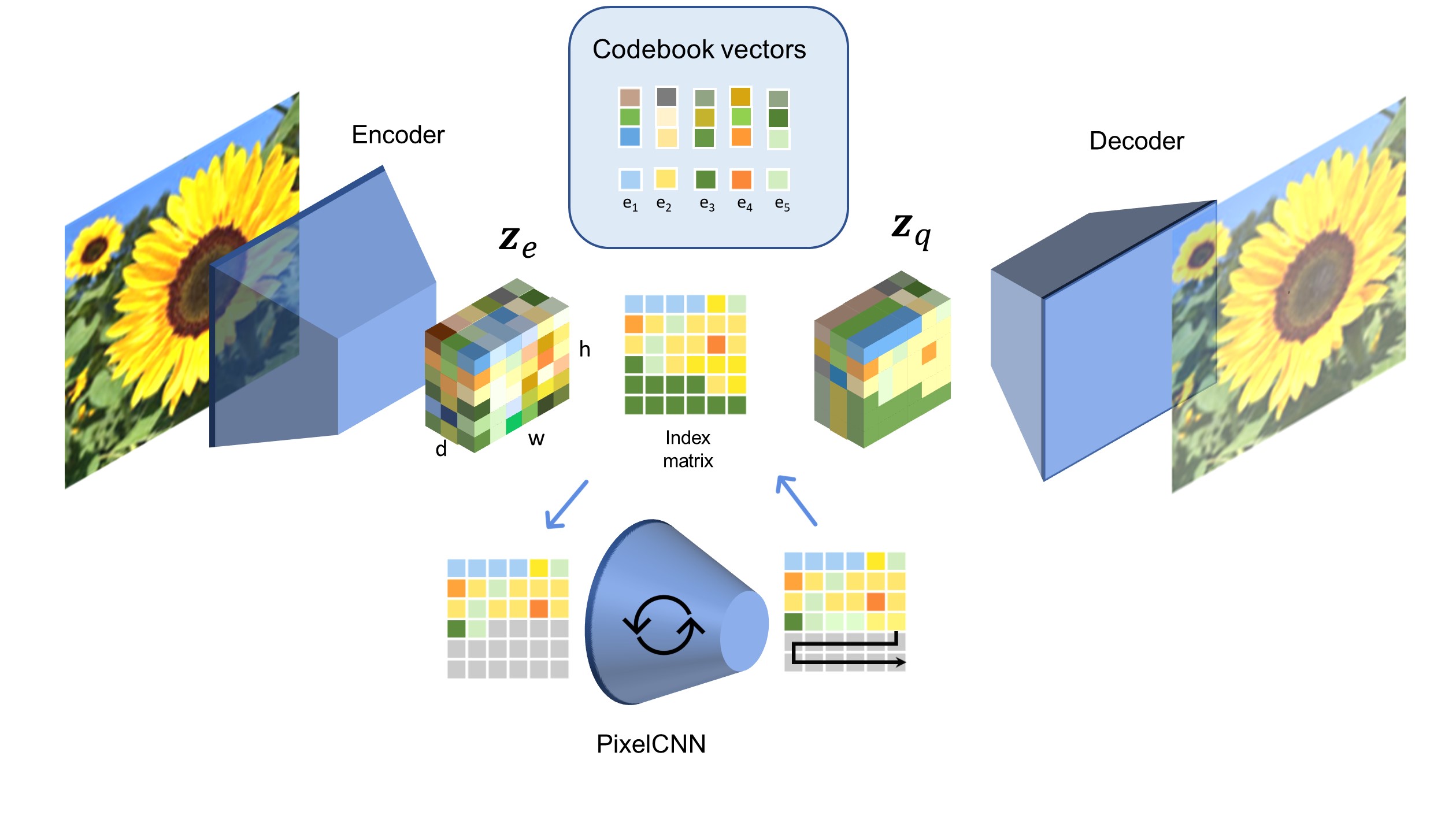}
\caption{{\bf Network architecture.}  The encoder, which consists of several
  convolutional layers, compresses the input image into an array $\mathbf{z}_e$ of
  $w \times h$ $d$-dimensional feature vectors. Each feature vector is then assigned to
  the closest codebook vector $\mathbf{e}_l$ to create the index matrix $z_x$ containing
  the indices $l$ and an array $\mathbf{z}_q$ of $w \times h$ corresponding
  $d$-dimensional codebook vectors $\mathbf{e}_l$. The decoder then reconstructs the
  original input based on the quantized array $\mathbf{z}_q$.  Attention is modeled by
  discarding consecutive entries in the lower part of the index matrix.  The missing part
  can be filled in by the PixelCNN in a recurrent process that performs semantic
  completion.  The completed index matrix is then passed to the decoder for
  reconstruction.}
\label{fig:vqvae}
\label{fig1}
\end{figure}

Our computational model consists of two networks known from the field of machine
learning, a Vector-Quantized Variational AutoEncoder (VQ-VAE) \cite{oord2017vqvae} and a
Pixel Convolutional Neural Network (PixelCNN) \cite{oord2016pixelcnn}, see
Figure~\ref{fig:vqvae}.

A VQ-VAE consists of an encoder, a decoder, and a latent representation between these
two. To quantize the latent representation there is also a set of codebook vectors, which
are optimized by vector quantization but otherwise fixed.  The VQ-VAE processes an input
image in the following steps, cf.\ Figure~\ref{fig:compframework}:
\begin{figure}
    \centering
    \includegraphics[width=\textwidth]{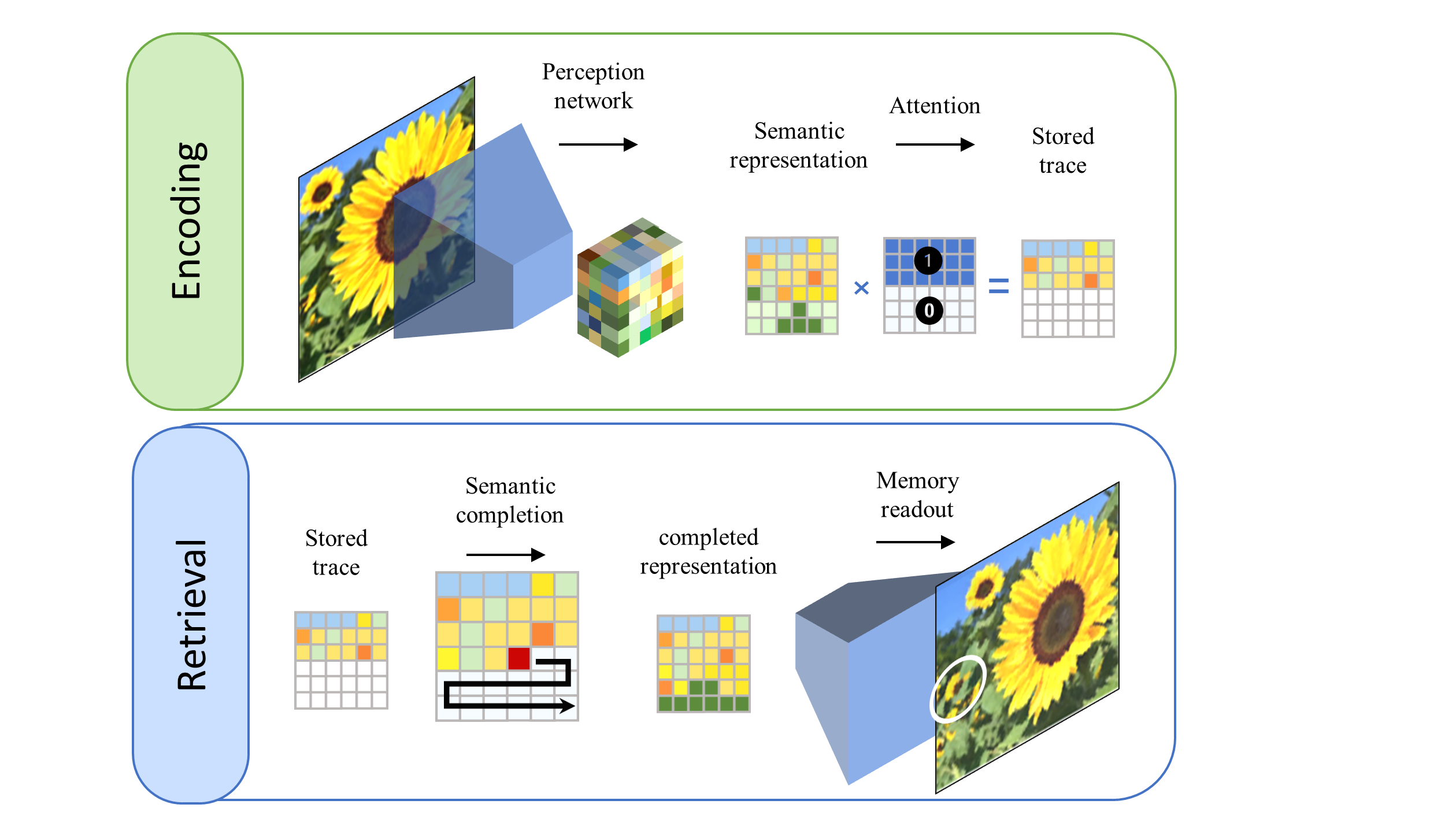}
    \caption{\textbf{Encoding and retrieval in the model.} Top: During encoding, the
      image is first passed though the encoder of the VQ-VAE and quantized to generate a
      semantic representation in form of the index matrix.  Attention then selects a
      fraction in the upper part, resulting in what is stored as a memory trace. Bottom:
      The stored memory trace is recalled and semantic completion by the PixelCNN fills
      in missing information.  The completed semantic information is then translated back
      from indices to codebook vectors and passed through the decoder of the VA-VAE to
      produce an image as a readout of the neural representation.  }
    \label{fig:compframework}
\end{figure}
\begin{enumerate}
\item\label{VQVAE:featureVectors} The VQ-VAE first compresses the image of size
  $28 \times 28$ (or $32 \times 32$ for images with context) with a convolutional neural
  network (the encoder) down to an array of $w\times h$ $d$-dimensional feature vectors
  ($w=h=7$ for the original images and $w=h=8$ for images with context, $d$ is set to
  $64$).  The positions within the array correspond to a grid of subsampled locations in
  the image.  Thus, this array still has a coarse spatial resolution, and the feature
  vectors are a description of the image around these locations.
\item This array is then converted to an array of $w\times h$ indices (called index
  matrix for short), each index indicating the $d$-dimensional codebook vector most
  similar to the feature vector at that position. This can be viewed as a quantization
  step that makes the representation more categorical, i.e.\ more semantic.  Up to this
  point the network has compressed the image into a more abstract and semantic
  representation.
\item In order to recover an image back from this highly compressed latent
  representation, the array of indices is converted to an array of the corresponding
  codebook vectors, which should be similar to the array of feature vectors of
  Step~\ref{VQVAE:featureVectors}.
\item The array of codebook vectors is then decompressed by a deconvolution neural
  network (the decoder mirroring the encoder) to produce an image of original size.
\end{enumerate}
A VQ-VAE alone can convert an image into a more abstract and semantic representation and
back again, but it is not generative in the sense that it could produce new reasonable
images from scratch or complement incomplete images.  This is fine as long as the full
index matrix is available.  However, we also want to model attentional selection, so that
only part of the index matrix can be recalled from memory.  In such cases we need a
generative component that is able to fill in the missing indices.  For that we use a
PixelCNN.

A PixelCNN is a probabilistic autoregressive generative model that is able to continue
sequences of numbers. It can fill in missing pixel RGB-values in an image in a fixed
sequence, for instance row-wise from top left to bottom right. If the first lines of an
image are given, the PixelCNN can continue the sequence of pixel RGB-values and fill in
the rest of the image reasonably well.  Since the PixelCNN is designed to continue
sequences and not to fill in missing pixels, it can only work in the one order it has
been trained for, here from top left to bottom right.  It cannot take advantage of known
pixel values in the bottom right corner for other pixels in the image.  Completing an
image with a PixelCNN is a very time consuming process. We apply the PixelCNN not to the
image but to the latent index matrix, which is much faster, to train the network and to
apply the trained network.  Since a PixelCNN only works in one particular order, we can
model attentional selection only in a primitive form by keeping the upper rows and
neglecting the lower rows of the index matrix.  The level of attention determines how
many rows to keep.

The VQ-VAE as well as the PixelCNN are both trained on a large set of training images.
First the VQ-VAE is trained to reconstruct the input images as well as possible, despite
the strong compression in the latent representation.  The weights of the encoder and
decoder are optimized as well as the codebook vectors.  Once the VQ-VAE is trained, the
PixelCNN can be trained on the index matrices generated by the trained VQ-VAE from the
training image.  See the Methods section for further details on the VQ-VAE and the
PixelCNN.

Our model is designed to reflect our hypothesis on generative episodic memory. That is,
the stored gist has far less information content than the input images; nonetheless, the
input can be reconstructed from it. The model captures complex statistics from the input
and also reflects the constructive nature of episodic memory that has been observed in
many studies.  When the attention is low (only a small part of the index matrix is
stored), the recalled memories are not necessarily faithful.  Still, they are valid and
likely reconstructions of the original data. The model is also capable of dreaming, i.e.,
generating unseen but valid episodes.

\subsection*{Analogy to the brain}

Even though VQ-VAE and PixelCNN originate from the field of machine learning, we believe
they capture essential aspects of neural processing in the brain, and they are an
appropriate level of description for our purposes here.

The encoder network of the VQ-VAE corresponds to the feedforward processing in the visual
system, which results in abstract object representations in the inferior temporal (IT)
cortex. Many studies have discussed a correspondence between the hierarchy of the human
visual areas and layers of CNNs \cite{kuzovkin2018activations,
  lindsay2021convolutional,yamins2014performance}. The decoder has a structure
symmetrical to the encoder and is similar to the feedback connections from higher levels
of the visual system to lower ones. It has been shown that during retrieval, a cortical
representation of the memory is formed in the lower levels of the visual pathway through
the feedback connections \cite{xia2015multilayered,takeda2019brain}.  Some studies have
actually used an autoencoder structure to model the feedback connections in the visual
pathway \cite{al2021reconstructing}.  In our model, the decoder generates a cortical
representation of the memory in its layers down to image level, which we take as a
readout of the cortical representation of the memory during retrieval.  A body of
research also indicates that there is semantic learning at the level of the visual
system, reflected in our model by the whole VQ-VAE network\cite{Hu2021SemanticIO}.

The PixelCNN learns statistical relationships between the elements of the latent
representation of the VQ-VAE by repeated exposure, i.e.\ it learns semantic information
from episodes akin to how it is hypothesized also for the brain
\cite{Michaelian2011-MICGM}.  It is then able to fill in missing elements in the semantic
representation of an image. We hypothesize that this is akin to a recurrent dynamics in
the higher cortical areas that can fill in missing information in a semantically
consistent and expected way \cite{tang2018recurrent,carrillo2020playing}.

We do not model storage in and retrieval from the hippocampus explicitly, we simply store
and recall a perfect copy of the selected parts of the index matrix, which represents the
episodic gist.  Storing just the indices of the codebook vectors, and not the vectors
themselves, is consistent with the indexing theory of hippocampal
memory\cite{Teyler1986TheHM}, although we would argue that our indices are also
represented in the cortex, so that semantic completion can take place there.

\section*{Results} 

Our model is able to process real-world images and we believe that sufficiently rich
statistical structure in the input patterns is essential for a meaningful simulation of
episodic memory. However, large images require large data sets and are computationally
expensive to process. As a compromise we use the well known MNIST data set of hand
written digits \cite{lecun1998mnist}, which is real-world, has a clear structure of ten
digit classes 0~to~9, and is of moderate size, so that simulations can be conducted
efficiently.  The images show white digits on black background with gray values
normalized between 0~and~1 and $28 \times 28$ pixels.  First we illustrate the behavior
of the system and then model a concrete memory task and compare it with experimental
results \cite{zoellner2021toaster}.

\subsection*{Scenario construction by semantic completion}

At the core of our model is the concept of an episodic gist, which is incomplete but can
be complemented by semantic information to reconstruct a full scenario from a partial
memory trace. What is being stored in the memory trace, i.e.\ what makes up the episodic
gist, is largely determined by attention. In our model, attentional control is somewhat
constrained and only determines how many consecutive indices of codebook vectors are
stored row-wise starting in the upper left corner. For low attention only the upper two
out of eight rows might be stored; for high attention the upper six rows might be
stored. The remaining part, if needed, has to be constructed based on semantic
information. It is important to note that attentional selection does not apply to the
images but to the latent representation in form of the index matrix.

\begin{figure}[htb!]
    \centering
    \includegraphics[width=\textwidth]{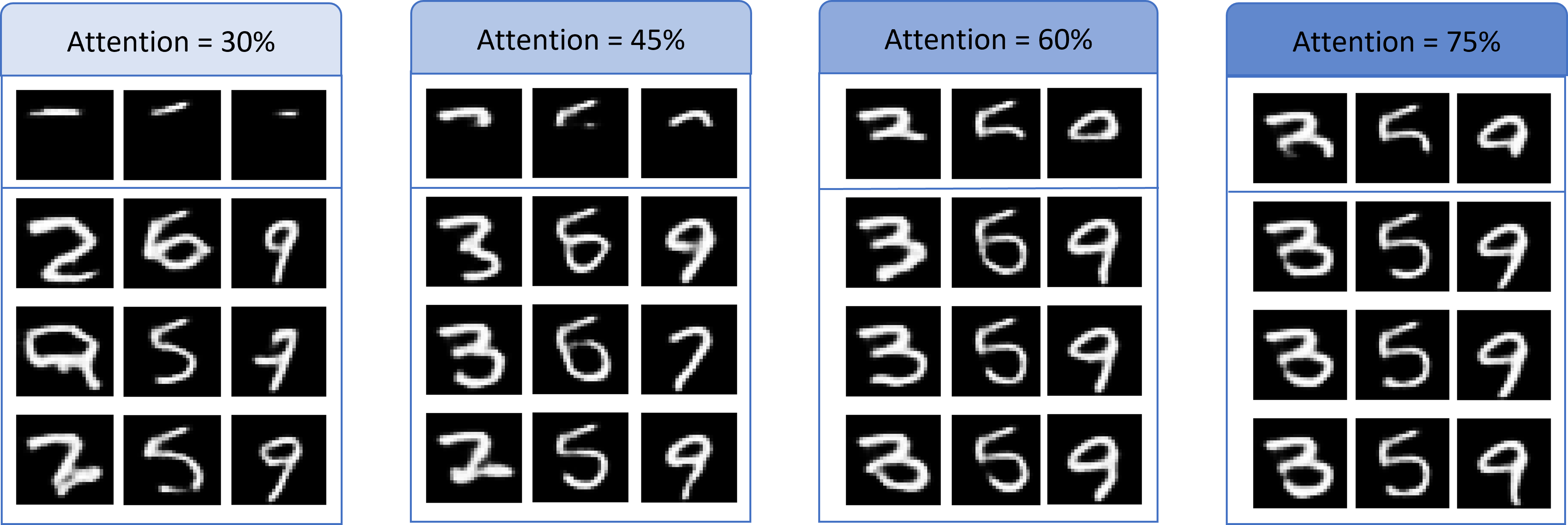}
    \caption{\textbf{Semantic completion.} Incomplete latent representations in the form
      of the index matrix are completed by the PixelCNN through a process we call
      semantic completion. The index matrix is already 30 times smaller than the image
      and only part of that representation is selected by attention. The attention level
      is defined as the percentage of the index matrix that is stored in the memory
      trace. The partial patterns in the first row visualize the episodic memory
      traces. The next three rows visualize three instances of the semantic completion
      based on the incomplete memory traces.}
    \label{fig:attention}
\end{figure}

To illustrate the effect of semantic completion we have compared recalled images from
memory traces at several different attention levels, see Figure~\ref{fig:attention}.
The first line is the reconstruction without semantic completion and the next lines are
three different reconstructions with semantic completion. At high attention, the
reconstruction is faithful; at low attention, the reconstruction is not necessarily
faithful but it is plausible given the attended parts.

\subsection*{Improved memory efficiency by semantic completion}

As mentioned earlier, the memory efficiency in our model is due to two factors: (i)~the
compression by the encoder network of the VQ-VAE and (ii)~the semantic completion by the
PixelCNN. The former reduces the storage requirement in our model already by a factor of
30 with little loss of image information, cf.\cite{walker2021predicting}.  Here we want
to address the latter factor.  To do so, we have to measure recall performance.  Even
though the mean squared distance between original image and recalled image is an obvious
and frequently used measure, it is not very useful as a measure of perceptual
similarity\cite{mathieu2015deep}.  We have therefore trained a classification network
(see the Methods section) to recognize the ten digits and evaluate the quality of recall
by the \emph{classification accuracy}, i.e. the percentage of correctly recognized
recalled patterns.

\begin{figure}
    \centering
    \includegraphics[width=0.49\textwidth]{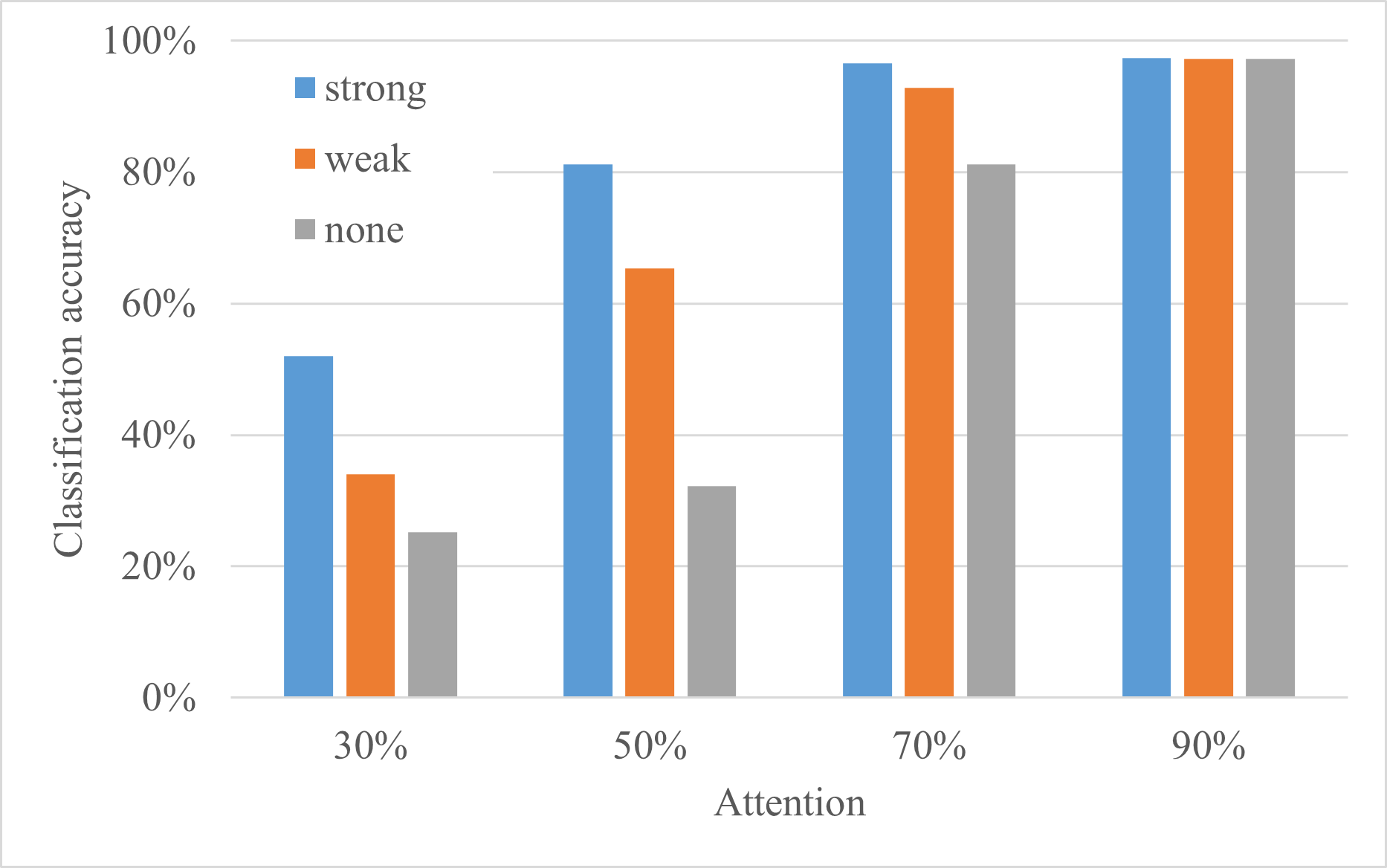}
    \includegraphics[width=0.49\textwidth]{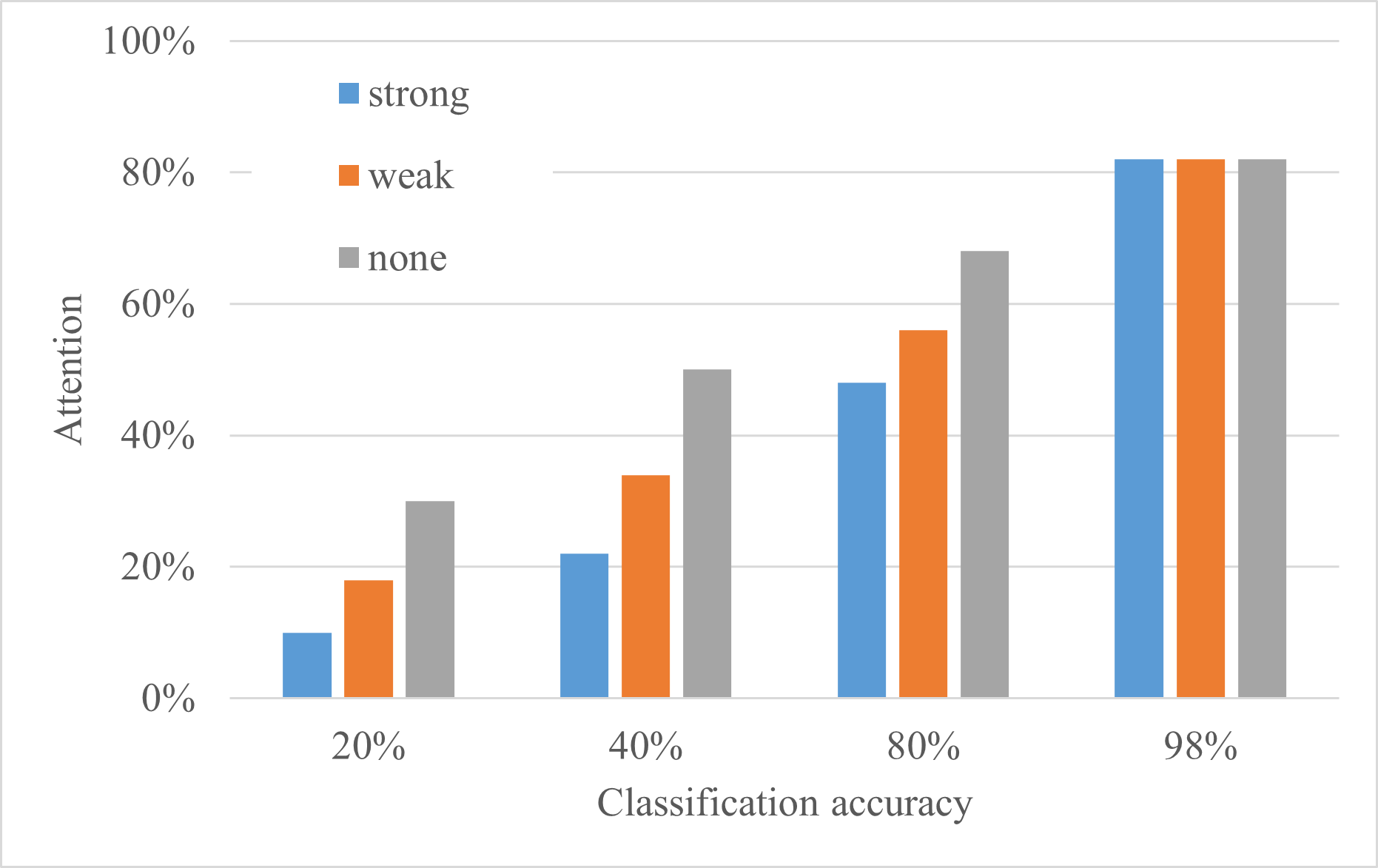}    
    \caption{\textbf{Semantic memory helps to increase the classification accuracy and
        capacity efficiency.} Left: A classifier trained on MNIST is used to evaluate the
      classification accuracy on recalled patterns of different attention levels. The
      stronger the PixelCNN with its semantic information the better the accuracy for any
      given attention level. Right: For any expected classification accuracy the stronger
      PixelCNN requires less attention.}
    \label{fig:semantic}
\end{figure}

Figure~\ref{fig:semantic} shows how semantic completion can improve memory efficiency.
We have run simulations with semantic completion by a fully trained PixelCNN, by a
partially trained PixelCNN, and without semantic completion at all, which we refer to as
\emph{strong}, \emph{weak}, and \emph{none} (semantic completion), respectively. The
strong network was trained for 40~epochs and had a loss of 0.65 (categorical cross
entropy), the weak network was trained for 3~epochs and had a loss of 0.75, and for the
none case the not attended parts were just filled black. The left panel shows the
improvement of the classification accuracy by semantic completion for fixed attention
levels; the right panel shows the saving in attention level by semantic completion for
fixed classification accuracy levels. We see that semantic completion can significantly
contribute to recall quality (left panel) or capacity (right panel) although the memory
saving is not nearly as large as for the compression by the encoder, maybe a factor of
two.

\subsection*{Semantic learning at the level of the VQ-VAE}

Semantic learning within the VQ-VAE has two aspects: the semantic learning within the
neural representations of the encoder and decoder used for efficient compression and the
categorization of the feature vectors by vector quantization.

\begin{figure}[!htb]
    \centering
    \includegraphics[width=\textwidth]{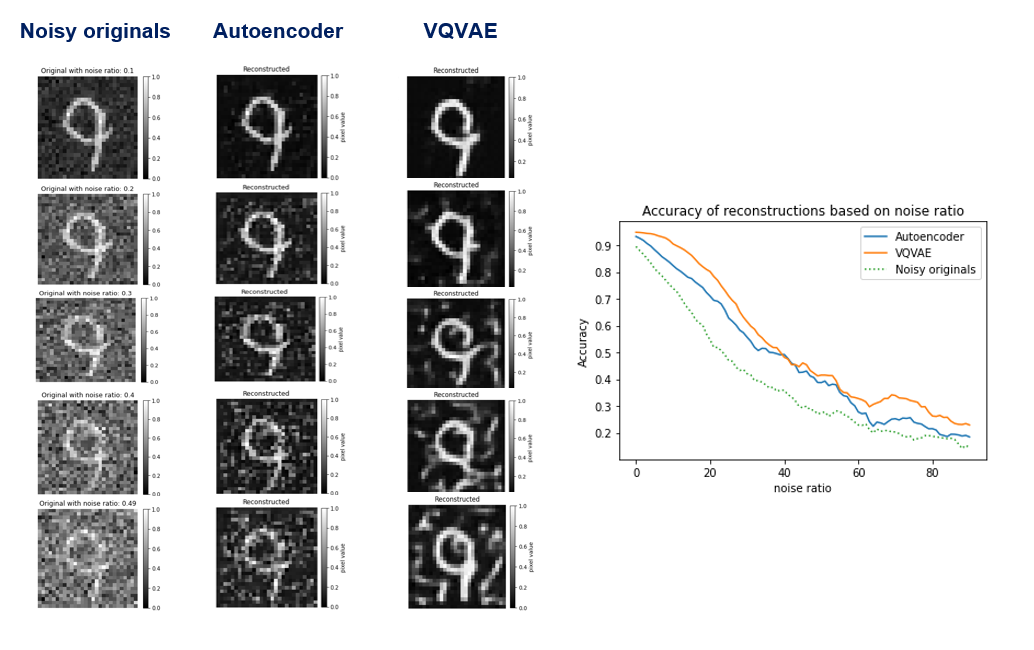}
    \caption{\textbf{Effect of autoencoding and quantization on noise reduction and
        classification performance.} Test images with different noise levels (left
      column) are processed by a VQ-VAE without (middle column) and with (right column)
      vector quantization.  An MNIST classifier is then used to determine the
      classification accuracy of the reconstructed images. The right panel shows the
      classification accuracy for the three settings against input noise ratio. The
      VQ-VAE network achieves highest classification accuracy for its reconstructed noisy
      inputs.}
    \label{fig:AE-VQVAE-noise}
\end{figure}

Figure~\ref{fig:AE-VQVAE-noise} shows how the semantic learning contributes to robustness
to noise.  Classification performance is compromised by pixel noise.  The
compression-decompression performed by just the autoencoder part of the VQ-VAE,
i.e. without the quantization, already reduces noise and improves performance;
quantization improves performance even further.

The noise reducing effect of autoencoders is well known\cite{Bhowick2019StackedAB}. The
quantization process in addition drags the internal representations towards what is know,
what it has seen before, and this helps it in becoming more robust to noise, because
noise moves the latent representation away from the normal distribution, while
quantization drags it back, imposing a denoising effect.

\begin{figure}[!htb]
    \centering
    \includegraphics[width=0.6\textwidth]{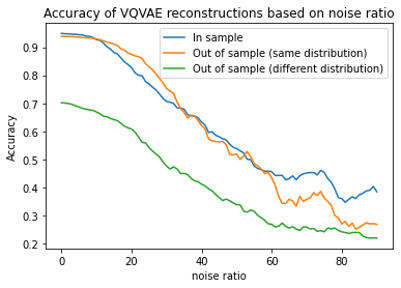}
    \caption{\textbf{Generalization of the VQ-VAE to other data sets.} A VQ-VAE network
      trained on half of the MNIST dataset digits (digits 0 to 4) is applied to three
      types of samples: MNIST images of digits 0~to~4 (in sample), MNIST images of digits
      5 to 9 (out of sample but share the same input distribution), and images from the
      Fashion MNIST dataset (out of sample and different than input distribution).  The
      reconstructed images of these three types of inputs are then tested for recognition
      accuracy using an MNIST classifier for digits 0~to~9 and a Fashion MNIST classifier
      images of clothes.}
    \label{fig:Generalization-aspect}
\end{figure}

\begin{figure}[!htb]
    \centering
    \includegraphics[width=0.6\textwidth]{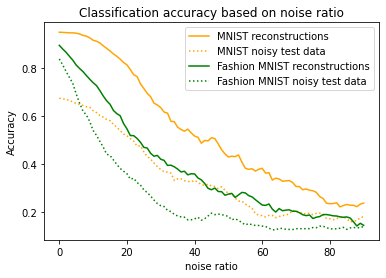}
    \caption{\textbf{Semantics learned by the VQ-VAE improves classification accuracy for
        out of distribution samples.}  The VQ-VAE has been trained on the MNIST data set,
      digits 0~to~9, and achieves an improvement in classification accuracy on noisy
      images also for the quite different stimuli of the Fashion MNIST data set.  }
    \label{fig:VQVAE-semantic}
\end{figure}

The latent representation in a VQ-VAE still has some spatial resolution and can take
advantage of the combinatorics in the index matrix to generalize to images from a
different distribution than the one trained on.  We have trained a VQ-VAE on MNIST digits
0~to~4 and then tested it on digits 5~to~9 as well as on the Fashion MNIST
dataset. Figure~\ref{fig:Generalization-aspect} shows that the VQ-VAE generalizes almost
perfectly from the low digits to the high ones, and it also generalizes to the quite
different fashion images of clothes to some degree. A VAE, for instance, would not be
able to that.
Furthermore, Figure~\ref{fig:VQVAE-semantic} shows that also the positive effect of the
VQ-VAE for robustness to noise illustrated in Figure~\ref{fig:AE-VQVAE-noise} generalizes
to the Fashion MNIST data base.

\subsection*{Modeling an episodic memory task}

An important goal of our modeling effort is to reproduce experimental results from
episodic memory research and eventually make suggestions and predictions for new
experiments. Here we relate to a recent experiment by Z\"{o}llner et al
\cite{zoellner2021toaster}.  The experiment takes place in three days. At day one,
participants are asked to move around an apartment in a virtual environment. The
apartment has three main rooms: bedroom, kitchen, and bathroom. Half of the objects in
the apartment are placed in a room where one would expect such an object, like a
microwave in the kitchen, this is referred to as \emph{congruent context}.  Half of them
are placed in a different room, like a toaster in a bathroom, referred to as
\emph{incongruent context}. Participants should first familiarize themselves with the
apartment and are then instructed to do some tasks and interact with some of both
congruent and incongruent objects, for example to make a sandwich with the toaster. This
provides some control over the level of attention with which the different objects are
perceived.  Participants are then tested on the next day on two tasks. In the recognition
task, participants rank how confident they are that they have seen a specific household
object and, if they think they have seen it, decide which room it was in. In the spatial
recall task, participants drag a household object into a map of the apartment and drop it
at the specific location where they think they have seen it. Both tasks are repeated
several times with different objects.  The same tasks are performed after seven days to
check how memory changes over time. Since the results show no significant difference
between day two and day eight and we are not modeling memory accuracy over time we pool
the data from both days.

\begin{figure}
    \centering
    \includegraphics[width=0.6\textwidth]{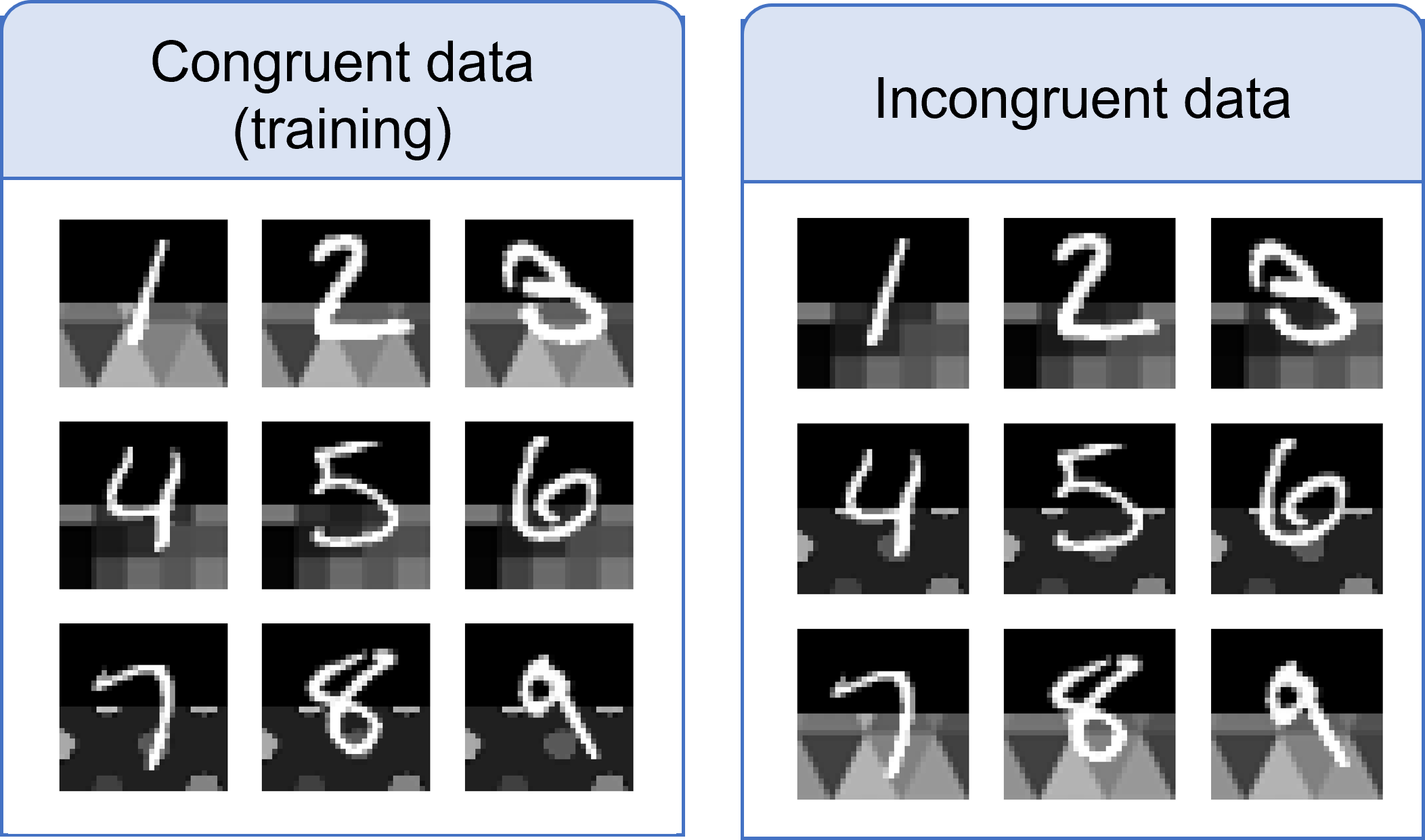}
    \caption{\textbf{Data with context as a background.} The algorithm is trained on the
      congruent data and learns the relationship between digits and the background in an
      unsupervised manner. Then it is tested on both the congruent and incongruent data.}
    \label{fig:data}
\end{figure}

To reproduce this experiment with our model we pad the images to size $32 \times 32$ and
augment them by three different backgrounds in the bottom half: A background with
triangles for digits 0~to~3, squares for digits 4~to~6, and circles for digits 7~to~9.
The backgrounds provide a context for the digits that can be congruent, e.g.\ a 3 in
front of triangles, or incongruent, e.g.\ a 5 in front of circles, see
Figure~\ref{fig:data}.  We use only the bottom half, so that it is possible to show the
model only the object without background, i.e.\ only the upper half.  With a more
flexible attention mechanism we could also use backgrounds across the whole image.
The model implicitly learns the association between a digit and its congruent context by
repeated exposure, e.g.\ 2 is always in the room with triangles.  The model is trained
only on congruent data and then tested on congruent as well as incongruent data.  The
different levels of attention are modeled by selecting various fractions of index matrix,
the latent representation.

The model simulation then goes as follows: First the VQ-VAE and the PixelCNN are trained
on the congruent data set.  Then a number of congruent and incongruent images are shown
to the system and stored in memory traces with varying levels of attention, i.e. with 5\%
(low attention), 52\% (medium attention), or 63\% (high attention) of the index matrix
stored.  The stored memory traces are then recalled by the network with semantic
completion by the PixelCNN.  A trained classifier for digits and another one for
backgrounds are used to model the responses of the humans.  If the digit classifier
recognizes the digit from the recalled image correctly, this counts as if the subject
remembers having seen the object.  Only then is the background classifier applied to
determine the type of background.

Since the PixelCNN has been trained only on congruent examples, it has a tendency to fill
in the congruent background in the bottom half of the image if it has stored a particular
number at attention level 52\%, i.e. only the top half. However, if attention level is
higher, say 63\%, the PixelCNN might infer an incongruent background from the bits that
are preserved about it in the memory trace and complete it. That means, for low attention
levels, one would expect that the model always recalls the congruent background, while
for high attention levels, it either recalls the episodically correct background or, if
it fails to do so, it recalls the congruent background.  Thus, the results should be
trivial for congruent images, because always the congruent background is recalled, but
interesting for incongruent images, because, depending on the attention level, the model
either correctly recalls the incongruent background or plausibly recalls the congruent
background.  It should usually not recall an incorrect and incongruent background.

\begin{figure}[htb!]
    \centering
    \includegraphics[width=0.485\textwidth]{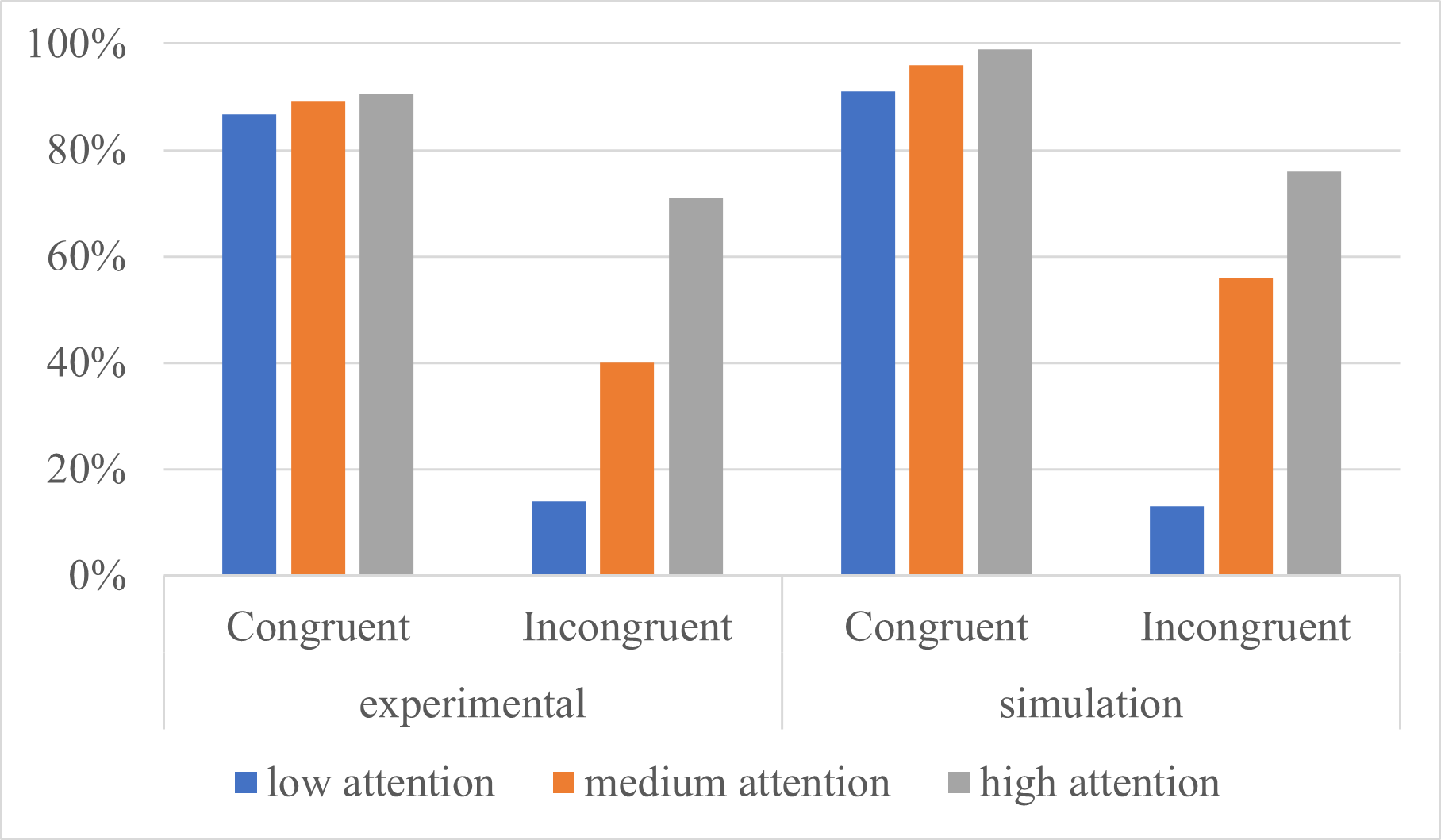}
    \includegraphics[width=0.45\textwidth]{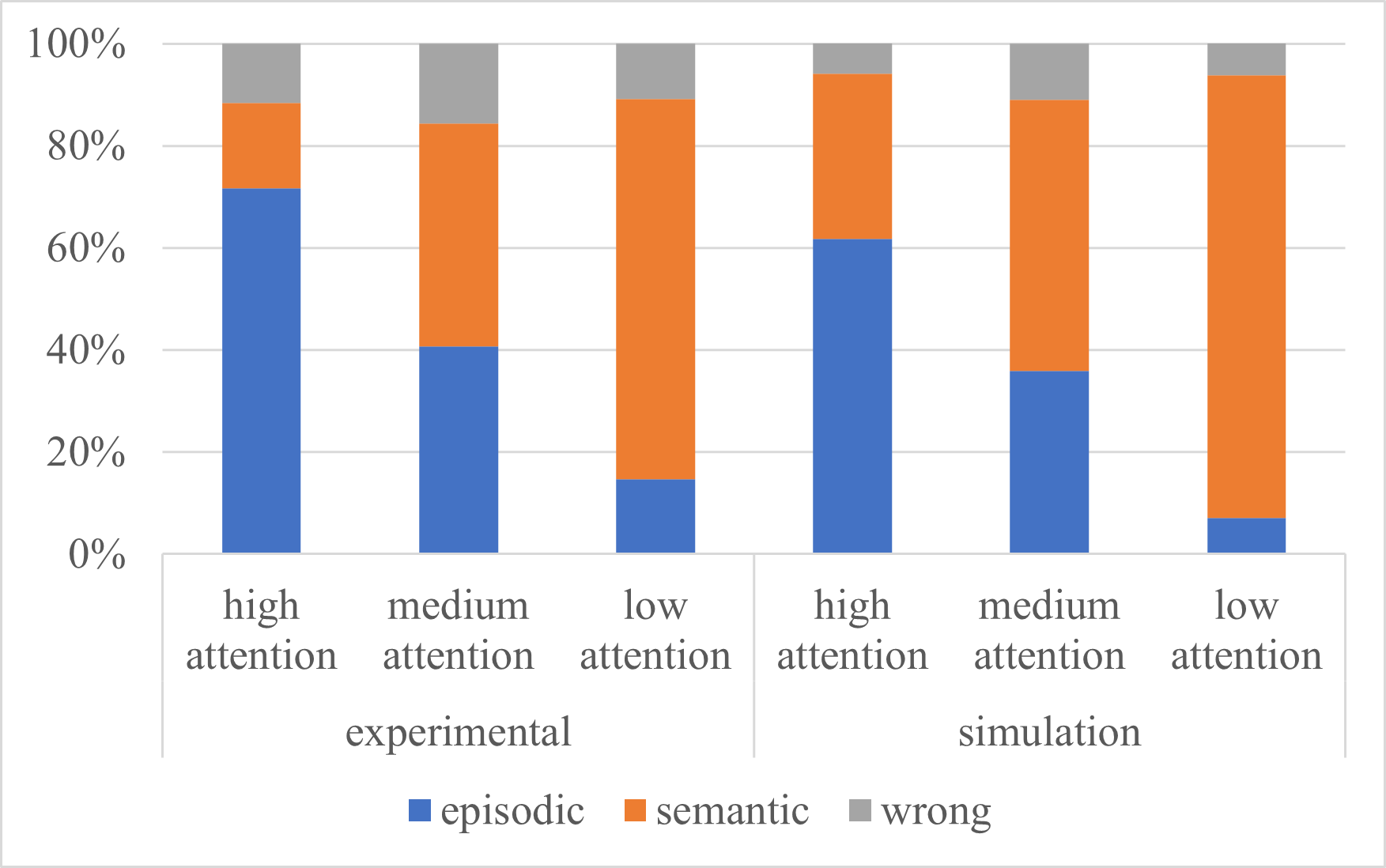}
    \caption{\textbf{The effect of context and attention in the model and experiment.}
      In the experiment, subjects explored an apartment in a virtual reality
      environment. Half of the household objects in the apartment were placed in an
      incongruent room. Later they were asked how sure they are if they have seen the
      incongruent objects and in which room they saw it. It is assumed that the objects
      that were reported to be seen with higher confidence had been seen with higher
      attention during the task. \emph{Episodic} means remembered correctly,
      \emph{semantic} means remembered in the semantically congruent room even though it
      was in an incongruent room and \emph{wrong} means remembered in an incongruent
      third room. The model matches the trend in the experimental data. Objects are
      remembered better in congruent context and attention improves the memory
      accuracy. Also in the incongruent case the items that were not remembered
      correctly, were more often remembered semantically than wrong.}
    \label{fig:comparison}
\end{figure}

Experimental as well as simulation result are shown in Figure~\ref{fig:comparison} in a
direct comparison.
The left panel shows the fraction of correctly recalled contexts for congruent and
incongruent cases depending on the attention level.  In both, experiments and model
simulations, we see the trivial behavior of high correct recall for the congruent cases
and a strong dependence on attention level for the incongruent cases.  But congruent
cases are always recalled better than incongruent ones, even for high attention levels.
The right panel shows more detailed result for the incongruent cases. The blue sections
indicate the correct recalls, called \emph{episodic} because they need to be explicitly
remembered; the orange sections indicate the incorrect but congruent recalls, called
\emph{semantic} because they can be semantically inferred from the object or digit; the
gray sections indicate the completely wrong recall, neither correct not congruent. The
simulation results match the experimental results quite well. To obtain a good fit, we
tuned the three attention levels in the model, which are not well constrained by the
experimentally determined attention levels in the subjects.  However, the appearance of
wrong recalls and their ratio to the semantic and episodic recalls emerge from the model.

The results in the experiments as well as the model simulations can be summarizes as
follows:
\begin{itemize}
\item Congruent contexts are recalled better than incongruent ones, as there is no
  conflict between episodic memory and semantic information.
\item Interaction with objects (i.e.\ paying attention) increases memory accuracy in both
  congruent and incongruent cases.
\item Contexts that are not remembered episodically correctly are more often remembered
  semantically congruently than completely wrong.
\end{itemize}

\section*{Conclusion}

Here, we present a model of generative episodic memory on a rather abstract level with a
network architecture combining known methods from machine learning, namely VQ-VAE and
PixelCNN. It can process real images, includes the potential for spatial attentional
selection (although still in a very primitive form), can represent images quite different
from those trained on, and models extraction of semantic information by compression,
which relates to abstraction, and quantization, which relates to categorization.  The two
models most closely related to ours \cite{bates2020efficient,kingma2013vae} are based on
a VQA, which lacks some of the advantages of a VQ-VAE and can therefore neither represent
quite different images nor model spatial attentional selection in any way, nor does it
include categorization of input features.

It is debatable whether machine learning methods are a good language to model the brain
like we do here. However, convolutional neural networks as well as recurrent neural
networks, on which our model is based, where inspired by the brain and are remarkably
successful also in computational neuroscience \cite{kuzovkin2018activations,
  lindsay2021convolutional,
  yamins2014performance,savage2019ai,papadimitriou2020brain}. Furthermore, they offer the
advantage of efficiency, so that real world images can be processed.  This is an
important factor, since the frequently used artificial random stimuli in earlier modeling
studies lack the statistical structure and regularities that are essential in studying
the interaction between episodic memory and semantic information.  Without statistical
regularities that can be exploited, episodic memory can by design not be generative.

Our model shows that and how generative episodic memory can work in principle. It
supports our conceptual framework for human episodic memory: (i)~Sensory input (an image
in our case) is processed by a multi-layer perceptual-semantic network, e.g.\ the visual
system, to generate a hierarchical more abstract representation. (ii)~Some aspects of
this representation, the episodic gist, are selected, presumably by attention and
depending on many factors. (iii)~The selected parts are then stored in form of a memory
trace in hippocampus.  The trace contains pointers to the selected semantic elements in
the hierarchical representation.  (iv)~During recall, a memory trace is reactivated in
hippocampus. (v)~The pointers in the memory trace in turn reactivate the semantic
elements. (vi)~Semantic information is finally used to fill in missing parts in a dynamic
process.  The latter step makes episodic memory generative, and we call the recall
process scenario construction.

In our model, Step~(i) is realized by the encoder of the VQ-VAE plus the quantization of
the feature vectors, Step~(ii) by selecting some fraction of the index matrix, Step~(iii)
by storing the partial index matrix, Steps~(iv) and ~(v) by retrieving the partial index
matrix, and Step~(vi) by the PixelCNN and the decoder of the VQ-VAE.

The model shows good semantic completion capabilities. It is remarkable how well the
PixelCNN generates plausible complete images even from small fragments, where faithful
reconstructions are not possible, see Figure~\ref{fig:semantic}. The system is even able
to generate plausible images from nothing (not shown).

Two factors in the model contribute to its capacity efficiency, compression and
decompression in the VQ-VAE as well as the semantic completion in the PixelCNN. We have
estimated that compression contributes a factor of about~30 while one can infer from
Figure~\ref{fig:semantic} that the semantic completion only contributes a factor of up
to~2 to the overall compression of input images into the memory traces. This is in line
with our hypothesis that semantic completion does not primarily contribute to the
capacity efficiency but serves other purposes.  Firstly, it is plausible to assume that
encoding and storage of an episode is limited by the attentional bottleneck during the
experience of the episode. Semantic completion can help to fill in those missing parts
that we were just not able to attend to in the episodic situation.  Secondly, semantic
completion can help to generalize better. It has been hypothesized that the main purpose
of episodic memory is not to remember the past but to help us make decisions for the
future\cite{schacter2007cognitive,schacter2007ghosts,la2016role}. Thus, if our knowledge
about the world changes, maybe also our memories of the past should change to be
maximally useful to deal with the future. Semantic completion can do exactly that.

We have successfully modeled the episodic memory experiment by Z\"{o}llner et al.\
\cite{zoellner2021toaster}. Both, experiments and simulations, show that congruent
contexts are recalled better than incongruent ones and that incorrectly recalled contexts
in incongruent cases are more often remembered semantically correct than completely
wrong. By designing the stimuli appropriately and tuning the selection percentages for
the index matrix for the three attention levels we were able to achieve a very good fit
to the experimental data. Due to the available degrees of freedom to match the
experimental results, our model does not yet have strong explanatory power. However, it
is not obvious and not a result of tuning that wrong recalls emerge in the incongruent
cases and that the ratio to the semantically correct recalls match so well. These are
emergent properties of the model.

Overall we feel this model advances our understanding and sharpens our concepts of
generative episodic memory. However, there are many ways in which the model can and
should be developed further.  (i)~The attentional selection is rather limited. As one of
the next steps we plan to make it more flexible.  (ii)~Even though the encoder is
hierarchical, consistent with our conceptual framework, the index matrix, on which the
selection and semantic completion are done, is not.  We plan to adopt a hierarchical
version of the VQ-VAE \cite{razavi2019vqvae2} to be able to apply semantic completion to
a truly hierarchical representation.  (iii)~The storage process in the hippocampus is not
modeled at all.  We plan to address this by adding a model for one-shot storage of
pattern sequences in hippocamal memory \cite{MelchiorBayatiEtAl-2019}.  This would also
allow us to investigate storage of sequential episodes, not just snapshots.
Sequentiality has been claimed to be one essential characteristics of episodic memory
\cite{cheng2016episodic, cheng2013crisp}.  (iv)~Besides developing the model further, we
will also continue to focus on modeling experiments on human episodic memory to constrain
the model better and contribute to the design of new experiments.

\section*{Acknowledgements}

We are grateful for very fruitful discussions with the group by Oliver Wolf and their
permission to work with their data.  This work was supported by a grant from the German
Research Foundation (DFG), project number 419039588 - FOR 2812, P5 (L.W.).

\section*{Methods}
 
Vector-Quantized Variational AutoEncoders (VQ-VAE) are autoencoders with a discrete
latent representation that process input in three steps. (i)~The encoder, which is a
Convolutional Neural Network (CNN), compresses the input $\mathbf{x}$ to generate the
encoding $\mathbf{z}_e$, which is an array of $w \times h$ $d$-dimensional feature
vectors.  (ii)~These feature vectors are then quantized with the help of $k$ codebook
vectors $\mathbf{e}_l \in \mathbb{R}^d$ according to the Vector Quantization (VQ)
frame\-work in equation \eqref{eq:vq}, i.e. each one is mapped to the closest codebook
vector $\mathbf{e}_l$, and index matrix $\mathbf{z}_x$ contains the corresponding
indices.

\newcommand\norm[1]{\left\lVert#1\right\rVert}
\begin{equation}
\begin{split}
    z_x^{(i,j)} & = \text{argmin}_l \norm{\mathbf{z}_e^{(i,j)}-\mathbf{e}_l}_2\\
    & \mathbf{z}_q^{(i,j)} =\mathbf{e}_{z_x^{(i,j)}}
    \\
    \text{with } i \in \{1, \dots&,w\} , j \in \{1,\dots,h\}, l \in \{1,\dots,k\} 
\label{eq:vq}
\end{split}
\end{equation}
The codebook vectors are initialized randomly, but they get optimized during training
together with the encoder and decoder. The quantized version of $\mathbf{z}_e$ is denoted
$\mathbf{z}_q$ and is a $w\times h$ array of codebook vectors. (iii)~The decoder, which
is a deconvolutional neural network, then generates a reconstruction $\mathbf{y}$ of the
input $\mathbf{x}$ based on the given $\mathbf{z}_q$.

A VQ-VAE can be trained end to end based on the loss function
\begin{equation} 
    L=\log{p}\left(\mathbf{y}=\mathbf{x}\mid \mathbf{z}_q\left(\mathbf{x}\right)\right)+ \norm{\text{ sg}\left[\mathbf{z}_e\left(\mathbf{x}\right)\right]-\mathbf{e}}_2^2+\beta \norm{\mathbf{z}_e\left(\mathbf{\mathbf{x}}\right)-\text{sg}\left[\mathbf{e}\right]}_2^2
    \label{eq:loss}
\end{equation}
The first term is the reconstruction loss, which is the negative log-likelihood of the
decoder output $\mathbf{y}$ being equal to the input $\mathbf{x}$, given the quantized
latent representation $\mathbf{z}_q$; this term optimizes the decoder and the
encoder. The second term is the VQ objective, which optimizes the codebook
vectors. 
This term uses the $l_2$ norm to push the codebook vectors towards $\mathbf{z}_e$ and
minimize the quantization error. Since the codebook vectors $\mathbf{e}_l$ may not train
as fast as the encoder, it might happen that the encoder outputs grow arbitrarily
large. To make sure that the encoder commits to the codebook vector the third term is
added. Essentially it pushes the encoder outputs toward the codebook vectors.  This third
term is called the commitment loss and constrains the size of the encoder outputs
\cite{oord2017vqvae}.

Since the quantization is a discrete operation, it is not possible to calculate its
gradient for back propagation. Therefore, the stop gradient (sg) operator is introduced
here.  During the forward pass it works like an identity operator. During the backward
pass (backpropagation), the gradient $\mathrm{\nabla}_zL$ is passed directly from
$\mathbf{z}_q$ to $\mathbf{z}_e$. The second and the third term have identical values,
the second one updates the codebook $\mathbf{e}$ via quantization (i.e. due to non-zero
$\mathrm{\nabla}_zL$), while the third one only affects the encoder.

A VQ-VAE by itself is not a generative model. However, it's quantized index matrix
($\mathbf{z}_x$) makes it possible to sample new data with the help of a PixelCNN
\cite{oord2016pixelcnn}. A PixelCNN is a well known autoregressive model, which is mainly
used to generate new images given the training data distribution. The basic principle of
this model is that each pixel in an image has a probability distribution that depends on
all the pixels that came before it:
\begin{equation}
    p\left(\mathbf{x}\right)=\prod_{i=1}^{D}{p\left(x_i\middle| x_{<i}\right) \text{ where }  x_{\mathit{<i}}=\left[\mathit{x}_1, \ldots,\mathit{x}_{\mathit{i}-1}\right]}
    \label{eq:pixelcnn}
\end{equation}
PixelCNNs generate images pixel by pixel and in a sequence (from top left to bottom
right) conditioned on all previously sampled pixels.  This process is slow for large
images, however in our case we use the PixelCNN only on the index matrix $\mathbf{z}_x$,
which is much smaller.  After training, the model can either complete a partial index
matrix or even generate a new one from scratch based on the semantic information learned
from the training data.  This is then converted to an array $\mathbf{z}_q$ of codebook
vectors and passed to the decoder to generate the output.

In this work we use a gated PixelCNN \cite{oord2016pixelcnn} with the activation function 
\begin{equation}
    \mathbf{y} = \tanh(W_{k,f} * \mathbf{x}) \odot \sigma (W_{k,g} * \mathbf{x})
    \label{Eq:activation}
\end{equation}
instead of the more common rectified linear activation function. $\sigma$ is the sigmoid
function, $k$ is the number of the layer, $\odot$ is the element-wise product and $*$
represents the convolution operator.  These multiplicative units help the network to
model more complex functions \cite{oord2016pixelcnn}.

The digit classifier network used in this paper is a basic CNN with three convolutional
layers, which was trained on digits in both congruent and incongruent context. This
network has an accuracy of 99.1\% on test data. The context classifier is a simple
pattern matching algorithm that assigns the pattern class based on mean square error.

\bibliography{ref.bib}

\begin{thebibliography}{10}

\bibitem{clayton2007episodic}
Clayton NS, Salwiczek LH, Dickinson A.
\newblock Episodic memory.
\newblock Current Biology. 2007;17(6):189-91.

\bibitem{reisberg2013oxford}
Reisberg D.
\newblock The Oxford handbook of cognitive psychology.
\newblock Oxford University Press; 2013.

\bibitem{tulving1972organization}
Tulving E.
\newblock Organization of memory.
\newblock Episodic and semantic memory. 1972.

\bibitem{bartlett1995remembering}
Bartlett FC, Bartlett FC.
\newblock Remembering: A study in experimental and social psychology.
\newblock Cambridge University Press; 1995.

\bibitem{sachs1967recognition}
Sachs JS.
\newblock Recognition memory for syntactic and semantic aspects of connected
  discourse.
\newblock Perception \& Psychophysics. 1967;2(9):437-42.

\bibitem{deese1959prediction}
Deese J.
\newblock On the prediction of occurrence of particular verbal intrusions in
  immediate recall.
\newblock Journal of experimental psychology. 1959;58(1):17.

\bibitem{roediger1995creating}
Roediger HL, McDermott KB.
\newblock Creating false memories: Remembering words not presented in lists.
\newblock Journal of experimental psychology: Learning, Memory, and Cognition.
  1995;21(4):803.

\bibitem{greenberg2010interdependence}
Greenberg DL, Verfaellie M.
\newblock Interdependence of episodic and semantic memory: Evidence from
  neuropsychology.
\newblock Journal of the International Neuropsychological society.
  2010;16(5):748-53.

\bibitem{devitt2017episodic}
Devitt AL, Addis DR, Schacter DL.
\newblock Episodic and semantic content of memory and imagination: A multilevel
  analysis.
\newblock Memory \& Cognition. 2017;45(7):1078-94.

\bibitem{hirst2012remembering}
Hirst W, Echterhoff G.
\newblock Remembering in conversations: The social sharing and reshaping of
  memories.
\newblock Annual review of psychology. 2012;63:55-79.

\bibitem{deuker2013playing}
Deuker L, M{\"u}ller AR, Montag C, Markett S, Reuter M, Fell J, et~al.
\newblock Playing nice: a multi-methodological study on the effects of social
  conformity on memory.
\newblock Frontiers in Human Neuroscience. 2013;7:79.

\bibitem{axmacher2010natural}
Axmacher N, Do~Lam AT, Kessler H, Fell J.
\newblock Natural memory beyond the storage model: repression, trauma, and the
  construction of a personal past.
\newblock Frontiers in human neuroscience. 2010;4:211.

\bibitem{herten2017role}
Herten N, Otto T, Wolf OT.
\newblock The role of eye fixation in memory enhancement under stress--An eye
  tracking study.
\newblock Neurobiology of learning and memory. 2017;140:134-44.

\bibitem{Schacter2020MemoryAI}
Schacter DL, Addis DR.
\newblock Memory and Imagination: Perspectives on Constructive Episodic
  Simulation.
\newblock The Cambridge Handbook of the Imagination. 2020.

\bibitem{addis2020mental}
Addis DR.
\newblock Mental time travel? A neurocognitive model of event simulation.
\newblock Review of Philosophy and Psychology. 2020;11(2):233-59.

\bibitem{cheng2016episodic}
Cheng S, Werning M.
\newblock What is episodic memory if it is a natural kind?
\newblock Synthese. 2016;193(5):1345-85.

\bibitem{cheng2011structure}
Cheng S, Frank LM.
\newblock The structure of networks that produce the transformation from grid
  cells to place cells.
\newblock Neuroscience. 2011;197:293-306.

\bibitem{neher2015memory}
Neher T, Cheng S, Wiskott L.
\newblock Memory storage fidelity in the hippocampal circuit: the role of
  subregions and input statistics.
\newblock PLoS computational biology. 2015;11(5):e1004250.

\bibitem{rolls1995model}
Rolls ET.
\newblock A model of the operation of the hippocampus and entorhinal cortex in
  memory.
\newblock International Journal of Neural Systems. 1995;6:51-70.

\bibitem{jensen1996novel}
Jensen O, Lisman JE.
\newblock Novel lists of $7\pm2$ known items can be reliably stored in an
  oscillatory short-term memory network: interaction with long-term memory.
\newblock Learning \& Memory. 1996;3(2-3):257-63.

\bibitem{becker2005computational}
Becker S.
\newblock A computational principle for hippocampal learning and neurogenesis.
\newblock Hippocampus. 2005;15(6):722-38.

\bibitem{cheng2016dissociating}
Cheng S, Werning M, Suddendorf T.
\newblock Dissociating memory traces and scenario construction in mental time
  travel.
\newblock Neuroscience \& Biobehavioral Reviews. 2016;60:82-9.

\bibitem{bates2020efficient}
Bates CJ, Jacobs RA.
\newblock Efficient data compression in perception and perceptual memory.
\newblock Psychological review. 2020;127(5):891.

\bibitem{nagy2020optimal}
Nagy DG, T{\"o}r{\"o}k B, Orb{\'a}n G.
\newblock Optimal forgetting: Semantic compression of episodic memories.
\newblock PLoS Computational Biology. 2020;16(10):e1008367.

\bibitem{kingma2013vae}
Kingma DP, Welling M.
\newblock Auto-encoding variational bayes.
\newblock arXiv preprint arXiv:13126114. 2013.

\bibitem{oord2017vqvae}
Oord Avd, Vinyals O, Kavukcuoglu K.
\newblock Neural discrete representation learning.
\newblock arXiv preprint arXiv:171100937. 2017.

\bibitem{SCHACTER2011467}
Schacter DL, Guerin SA, Jacques PLS.
\newblock Memory distortion: an adaptive perspective.
\newblock Trends in Cognitive Sciences. 2011;15(10):467-74.

\bibitem{koutstaal1997gist}
Koutstaal W, Schacter DL.
\newblock Gist-based false recognition of pictures in older and younger adults.
\newblock Journal of memory and language. 1997;37(4):555-83.

\bibitem{oliva2005gist}
Oliva A.
\newblock Gist of the scene.
\newblock In: Neurobiology of attention. Elsevier; 2005. p. 251-6.

\bibitem{graham2000insights}
Graham KS, Simons JS, Pratt KH, Patterson K, Hodges JR.
\newblock Insights from semantic dementia on the relationship between episodic
  and semantic memory.
\newblock Neuropsychologia. 2000;38(3):313-24.

\bibitem{fang2018interaction}
Fang J, R{\"u}ther N, Bellebaum C, Wiskott L, Cheng S.
\newblock The interaction between semantic representation and episodic memory.
\newblock Neural Computation. 2018;30(2):293-332.

\bibitem{collins1969retrieval}
Collins AM, Quillian MR.
\newblock Retrieval time from semantic memory.
\newblock Journal of verbal learning and verbal behavior. 1969;8(2):240-7.

\bibitem{irish2013pivotal}
Irish M, Piguet O.
\newblock The pivotal role of semantic memory in remembering the past and
  imagining the future.
\newblock Frontiers in behavioral neuroscience. 2013;7:27.

\bibitem{oord2016pixelcnn}
Oord Avd, Kalchbrenner N, Vinyals O, Espeholt L, Graves A, Kavukcuoglu K.
\newblock Conditional image generation with pixelcnn decoders.
\newblock arXiv preprint arXiv:160605328. 2016.

\bibitem{kuzovkin2018activations}
Kuzovkin I, Vicente R, Petton M, Lachaux JP, Baciu M, Kahane P, et~al.
\newblock Activations of deep convolutional neural networks are aligned with
  gamma band activity of human visual cortex.
\newblock Communications biology. 2018;1(1):1-12.

\bibitem{lindsay2021convolutional}
Lindsay GW.
\newblock Convolutional neural networks as a model of the visual system: Past,
  present, and future.
\newblock Journal of cognitive neuroscience. 2021;33(10):2017-31.

\bibitem{yamins2014performance}
Yamins DL, Hong H, Cadieu CF, Solomon EA, Seibert D, DiCarlo JJ.
\newblock Performance-optimized hierarchical models predict neural responses in
  higher visual cortex.
\newblock Proceedings of the national academy of sciences.
  2014;111(23):8619-24.

\bibitem{xia2015multilayered}
Xia R, Guan S, Sheinberg DL.
\newblock A multilayered story of memory retrieval.
\newblock Neuron. 2015;86(3):610-2.

\bibitem{takeda2019brain}
Takeda M.
\newblock Brain mechanisms of visual long-term memory retrieval in primates.
\newblock Neuroscience research. 2019;142:7-15.

\bibitem{al2021reconstructing}
Al-Tahan H, Mohsenzadeh Y.
\newblock Reconstructing feedback representations in the ventral visual pathway
  with a generative adversarial autoencoder.
\newblock PLoS Computational Biology. 2021;17(3):e1008775.

\bibitem{Hu2021SemanticIO}
Hu R, Jacobs RA.
\newblock Semantic influence on visual working memory of object identity and
  location.
\newblock Cognition. 2021;217.

\bibitem{Michaelian2011-MICGM}
Michaelian K.
\newblock Generative Memory.
\newblock Philosophical Psychology. 2011;24(3):323-42.

\bibitem{tang2018recurrent}
Tang H, Schrimpf M, Lotter W, Moerman C, Paredes A, Caro JO, et~al.
\newblock Recurrent computations for visual pattern completion.
\newblock Proceedings of the National Academy of Sciences.
  2018;115(35):8835-40.

\bibitem{carrillo2020playing}
Carrillo-Reid L, Yuste R.
\newblock Playing the piano with the cortex: role of neuronal ensembles and
  pattern completion in perception and behavior.
\newblock Current opinion in neurobiology. 2020;64:89-95.

\bibitem{Teyler1986TheHM}
Teyler TJ, Discenna P.
\newblock The hippocampal memory indexing theory.
\newblock Behavioral neuroscience. 1986;100 2:147-54.

\bibitem{lecun1998mnist}
LeCun Y. The MNIST database of handwritten digits; 1998.
\newblock Available from: \url{http://yann.lecun.com/exdb/mnist/}.

\bibitem{zoellner2021toaster}
Z\"{o}llner C, Klein N, Cheng S, Schubotz R, Axmacher N, Wolf OT. Where was the
  Toaster? Interplay of Episodic Memory Traces and Semantic Knowledge during
  Scenario Construction. PsyArXiv; 2021.
\newblock Available from: \url{https://psyarxiv.com/2kmwy}.

\bibitem{walker2021predicting}
Walker J, Razavi A, Oord Avd.
\newblock Predicting Video with VQVAE.
\newblock arXiv preprint arXiv:210301950. 2021.

\bibitem{mathieu2015deep}
Mathieu M, Couprie C, LeCun Y.
\newblock Deep multi-scale video prediction beyond mean square error.
\newblock arXiv preprint arXiv:151105440. 2015.

\bibitem{Bhowick2019StackedAB}
Bhowick D, Gupta DK, Maiti S, Shankar U.
\newblock Stacked autoencoders based machine learning for noise reduction and
  signal reconstruction in geophysical data.
\newblock arXiv: Computational Engineering, Finance, and Science. 2019.

\bibitem{savage2019ai}
Savage N.
\newblock How AI and neuroscience drive each other forwards.
\newblock Nature. 2019;571(7766):S15-5.

\bibitem{papadimitriou2020brain}
Papadimitriou CH, Vempala SS, Mitropolsky D, Collins M, Maass W.
\newblock Brain computation by assemblies of neurons.
\newblock Proceedings of the National Academy of Sciences.
  2020;117(25):14464-72.

\bibitem{schacter2007cognitive}
Schacter DL, Addis DR.
\newblock The cognitive neuroscience of constructive memory: remembering the
  past and imagining the future.
\newblock Philosophical Transactions of the Royal Society B: Biological
  Sciences. 2007;362(1481):773-86.

\bibitem{schacter2007ghosts}
Schacter DL, Addis DR.
\newblock The ghosts of past and future.
\newblock Nature. 2007;445(7123):27-7.

\bibitem{la2016role}
La~Corte V, Piolino P.
\newblock On the Role of Personal semantic memory and temporal distance in
  episodic future thinking: the TEDIFT model.
\newblock Frontiers in human neuroscience. 2016;10:385.

\bibitem{razavi2019vqvae2}
Razavi A, van~den Oord A, Vinyals O. Generating Diverse High-Fidelity Images
  with VQ-VAE-2; 2019.

\bibitem{MelchiorBayatiEtAl-2019}
Melchior J, Bayati M, Azizi A, Cheng S, Wiskott L. A Hippocampus Model for
  Online One-Shot Storage of Pattern Sequences; 2019.
\newblock e-print arXiv:1905.12937.
\newblock Available from: \url{https://arxiv.org/abs/1905.12937}.

\bibitem{cheng2013crisp}
Cheng S.
\newblock The CRISP theory of hippocampal function in episodic memory.
\newblock Frontiers in neural circuits. 2013;7:88.

\end{thebibliography}

\end{document}